\documentclass[sigconf]{acmart}

\usepackage{booktabs} 
\usepackage{multirow}

\setcopyright{rightsretained}

\acmDOI{10.475/123_4}

\acmISBN{123-4567-24-567/08/06}

\acmConference[GLSVLSI'18]{ACM GLSVLSI conference}{May 2018}{Chicago, Illinois USA} 
\acmYear{1997}
\copyrightyear{2016}

\acmArticle{4}
\acmPrice{15.00}

\begin{document}

\copyrightyear{2018} 
\acmYear{2018} 
\setcopyright{licensedusgovmixed}
\acmConference[GLSVLSI '18]{2018 Great Lakes Symposium on VLSI}{May 23--25, 2018}{Chicago, IL, USA}
\acmBooktitle{GLSVLSI '18: 2018 Great Lakes Symposium on VLSI, May 23--25, 2018, Chicago, IL, USA}
\acmPrice{15.00}
\acmDOI{10.1145/3194554.3194558}
\acmISBN{978-1-4503-5724-1/18/05}
\pagestyle{empty}

\title{Low-Energy Deep Belief Networks Using Intrinsic Sigmoidal Spintronic-based Probabilistic Neurons}

\author{Ramtin Zand\textsuperscript{1}, Kerem Yunus Camsari\textsuperscript{2}, Steven D. Pyle\textsuperscript{1}, Ibrahim Ahmed\textsuperscript{3}, \\ Chris H. Kim\textsuperscript{3}, and Ronald F. DeMara\textsuperscript{1}}
\affiliation{%
  \institution{\textsuperscript{1}Department of Electrical and Computer Engineering, University of Central Florida, Orlando, FL, 32816 USA}
}
\affiliation{%
  \institution{\textsuperscript{2}Department of Electrical and Computer Engineering, Purdue University, West Lafayette, IN, 47906 USA}
}

\affiliation{%
  \institution{\textsuperscript{3}Department of Electrical and Computer Engineering, University of Minnesota, Minneapolis, MN 55455 USA}
}
\renewcommand{\shortauthors}{R. Zand et al.}

\begin{abstract}
A low-energy hardware implementation of deep belief network (DBN) architecture is developed using near-zero energy barrier probabilistic spin logic devices (p-bits), which are modeled to realize an intrinsic sigmoidal activation function. A CMOS/spin based weighted array structure is designed to implement a restricted Boltzmann machine (RBM). Device-level simulations based on precise physics relations are used to validate the sigmoidal relation between the output probability of a p-bit and its input currents. Characteristics of the resistive networks and p-bits are modeled in SPICE to perform a circuit-level simulation investigating the performance, area, and power consumption tradeoffs of the weighted array. In the application-level simulation, a DBN is implemented in MATLAB for digit recognition using the extracted device and circuit behavioral models. The MNIST data set is used to assess the accuracy of the DBN using 5,000 training images for five distinct network topologies. The results indicate that a baseline error rate of 36.8\% for a 784$\times$10 DBN trained by 100 samples can be reduced to only 3.7\% using a 784$\times$800$\times$800$\times$10 DBN trained by 5,000 input samples. Finally, Power dissipation and accuracy tradeoffs for probabilistic computing mechanisms using resistive devices are identified.  
\end{abstract}

%
%

\maketitle 

\section{Introduction}
The interrelated fields of machine learning (ML), and artificial neural networks (ANN) have grown significantly in previous decades due to the availability of powerful computing systems to train and simulate large scale ANNs within reasonable time-scales, as well as the abundance of data available to train such networks in recent years. The resulting research has realized a bevy of ANN architectures that have performed incredible feats including a wide range of classification problems, and various recognition tasks.

Most ML techniques in-use today rely on supervised learning, where the systems are trained on patterns with a known desired output, or label. However, intelligent biological systems exhibit unsupervised learning whereby statistically correlated input modalities are associated within an internal model used for probabilistic inference and decision making \cite{buesing2011}. One interesting class of unsupervised learning approaches that has been extensively researched is the Restricted Boltzmann machine (RBM) \cite{hinton2006}. RBMs can be hierarchically organized to realize deep belief networks (DBNs) that have demonstrated unsupervised learning abilities, such as natural language understanding \cite{Sarikaya2014}. Most RBM and DBN research has focused on software implementations, which provides flexibility, but requires significant execution time and energy due to large matrix multiplications that are relatively inefficient when implemented on standard Von-Neumann architectures due to the memory-processor bandwidth bottleneck when compared to hardware-based in-memory computing approaches \cite{Merolla2014}. Thus, research into hardware-based RBM designs has recently sought to alleviate these constraints. 

Previous hardware-based RBM implementations have aimed to overcome software limitations by utilizing FPGAs \cite{Kim2010,Ly2010} and stochastic CMOS \cite{Ardakani2017}. In recent years, emerging technologies such as resistive RAM (RRAM) \cite{SHERI2015,Bojnordi2016} and phase change memory (PCM) \cite{Eryilmaz2016} are proposed to be leveraged within the DBN architecture as weighted connections interconnecting building blocks in RBMs. While most of the previous hybrid Memristor/CMOS designs focus on improving the synapse behaviors, the work presented herein overcomes many of the preceding challenges by utilizing a novel spintronic p-bit device that leverages intrinsic thermal noise within low energy barrier nanomagnets to provide a natural building block for RBMs within a compact and low-energy package. The contribution of this paper is go to beyond using low-energy barrier magnetic tunnel junctions (MTJs), as has been previously introduced for a neuron in spiking neuromorphic systems \cite{sengupta2016magnetic,Sengupta2016prob}. To the best of our knowledge this paper is the first effort to use MTJs with near-zero energy barriers as neurons within an RBM implementation. Additionally, various parameters of a hybrid CMOS/spin weight array structure are investigated for metrics of power dissipation, and error rate using the MNIST digit recognition benchmarks.

\section{Fundamentals of RBM}
Boltzmann Machines (BM) are a class of recurrent stochastic ANNs with binary nodes whereby each possible state of the network, \textit{v}, has an energy determined by the undirected connection weights between nodes and the node bias as described by (1), where $s_i^v$ is the state of node \textit{i} in \textit{v}, \textit{b\textsubscript{i}} is the bias, or intrinsic excitability of node \textit{i}, and \textit{w\textsubscript{ij}} is the connection weight between nodes \textit{i} and \textit{j} \cite{ackley1985}.
\begin{equation}
  E(v) = -\sum_{i} s_i^v b_i -\sum_{i<j} s_i^v s_j^v w_{ij} 
\end{equation}

\begin{equation}
  P(s_i = 1) = \sigma (b_i + \sum_{j} w_{ij} s_j)
\end{equation}

Each node in a BM has a probability to be in state 1 according to (2), where $\sigma$ is the logistic sigmoid function. BMs, when given enough time, will reach a Boltzmann distribution where the probability of the system being in state \textit{\textbf{v}} is found by $P(v) = \frac{e^{-E(v)}}{\sum_{u} e^{-E(u)}}$, where \textit{\textbf{u}} could be any possible state of the system. Thus, the system is most likely to be found in states that have the lowest associated energy.
Restricted Boltzmann machines (RBMs) are BMs constrained to two fully-connected non-recurrent layers called the \textit{visible layer}, where salient inputs clamp nodes to output levels of either zero or one, and the \textit{hidden layer}, where associations between input vectors are learned. By enforcing the conditional independence of the visible and hidden layers, unbiased samples from the input can be obtained in one time-step, which enhances the learning process. 

The most widely used method for training RBMs is contrastive divergence (CD), which is an approximate gradient descent procedure using Gibbs sampling \cite{carreira2005}. CD operates in three phases: \textit{(1) Positive Phase:} A training input vector, \textbf{$v$}, is applied to the visible layer by clamping the nodes to either 1 or 0 levels, and the hidden layer is sampled, \textbf{$h$}. \textit{(2) Negative Phase:} by clamping the hidden layer to \textbf{$h$}, the reconstructed input layer is sampled, \textbf{$v'$}. Then, clamp the visible layer to v' and sample the hidden layer to obtain \textbf{$h'$}. \textit{(3)} Update the weights according to $\Delta W = \eta (vh^T-v'h'^T)$, where $\eta$ is the learning rate and \textbf{$W$} is the weight matrix. 

DBNs are realized when additional hidden layers are stacked on top of an RBM, and can be trained in a very similar way to RBMs. Essentially, training a DBN involves performing CD on the visible layer and the first hidden layer for as many steps as desired, then fixing those weights and moving up a hierarchy as follows. The first hidden layer is now viewed as a visible layer, while the second hidden layer acts as a hidden layer with respect to the CD procedure identified above. Next, another set of CD steps are performed, and then the process is repeated for each additional layer of the DBN.

\section{Spin-Based Building Block For RBM}
In this section, we provide a detailed description of the p-bit that provides the building block for our proposed spin-based BM architecture. Individual building blocks are interconnected by networks of memristive devices whose resistances can be programmed to provide the desired weights. For instance, in this paper, we will assume that the memristive devices are implemented using the three terminal spin-orbit torque (SOT)-driven domain wall motion (DWM) device proposed in \cite{Sengupta2016hybrid}.

\begin{figure}
\includegraphics[scale=0.12]{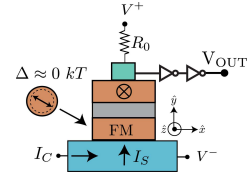}
\caption{Structure of a p-bit.}
\end{figure}

\begin{figure}
\includegraphics[scale=0.28]{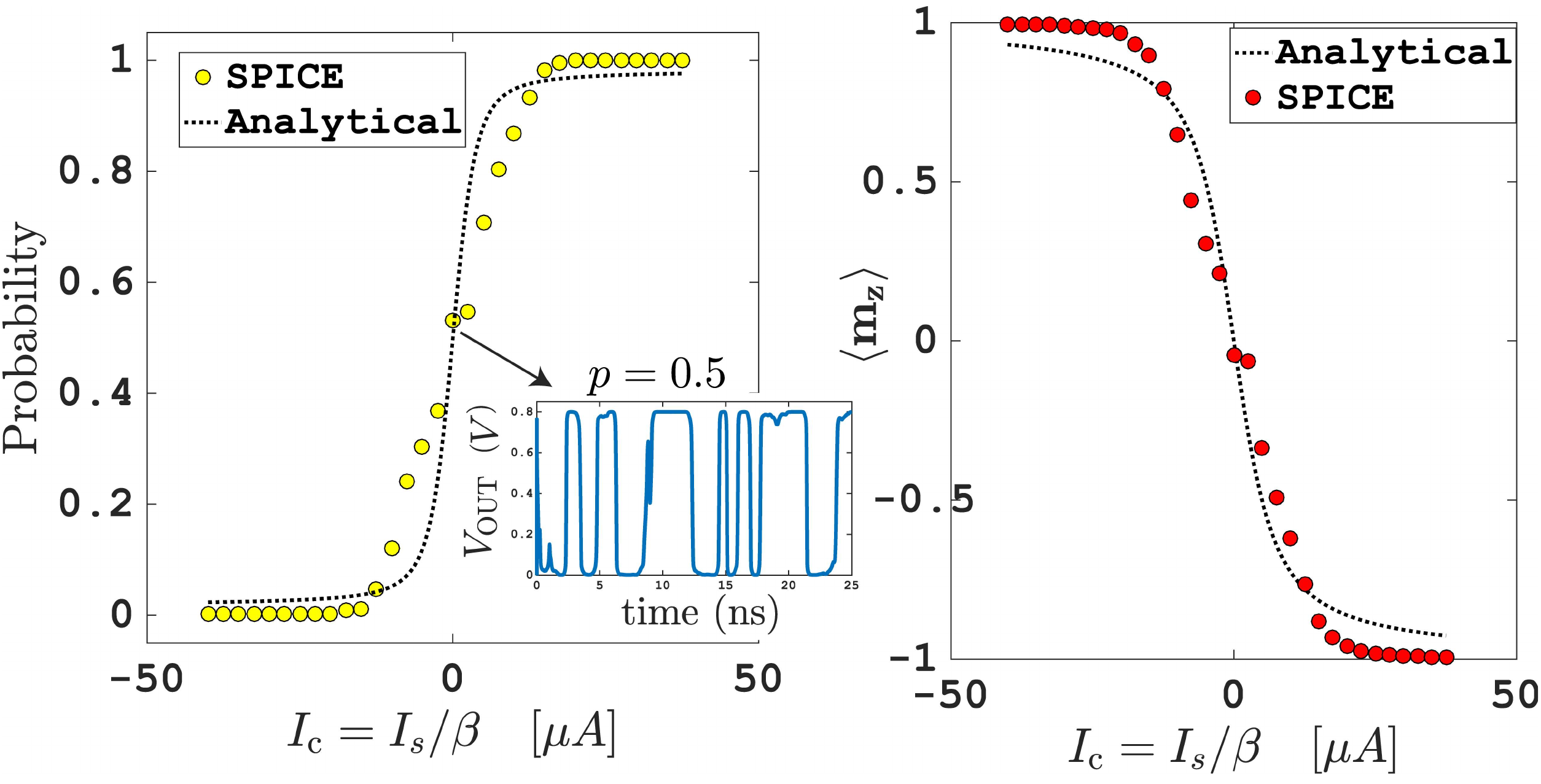}
\caption{Time-averaged results over 100 ns for p-bit.}
\end{figure}

The activation function is achieved by a spintronic building block that has been used in the design of probabilistic spin logic devices (p-bits) for a wide variety of Boolean and non-Boolean problems \cite{Camsari2017,Faria2017,sutton2017,behin2016}. The basic functionality of the p-bit shown in Fig. 1 \cite{Camsari2017} is to produce a stochastic output whose steady-state probability is modulated by an input current to generate a sigmoidal activation function. For instance, a high positive input current produces a stochastic output with a high probability of ``0'', and vice versa. In the absence of any input current, the device generates either 0 or VDD outputs with roughly equal probability of 0.5, as shown in Fig. 2. This device consists of a 3-terminal, spin-Hall driven MTJ \cite{Liu2012} that uses a circular, unstable nanomagnet ($\Delta \ll 40kT$), whereby its output is amplified by CMOS inverters as shown in Fig. 1. This MTJ with an unstable free layer can be fabricated using standard technology such that the surface anisotropy to achieve perpendicular magnetic anisotropy (PMA) that is not strong enough to overcome the demagnetizing field. Thus, the magnetization stochastically rotates in the plane, due to the presence of thermal fluctuations. 
 
The charge current that is injected to the spin-Hall layer creates a spin-current flowing into the circular FM (in the +y direction), which does not have an axis with any preferential geometry. The spin-polarization of this spin-current is in the ($\pm z$) direction, and pins the magnetization in the (+z) or (-z) direction depending on the direction of the charge current, through the spin-torque mechanism \cite{sutton2017}. The inherent physics of the spin-current driven low-barrier nanomagnet provides a natural sigmoidal function when a long time average of magnetization is taken. Through the tunneling magnetoresistance effect, a charge current flowing through the MTJ with a stable fixed layer detects the modulated magnetization as a voltage change. To achieve this, a small read voltage $V_R$ is applied between the $V+$ and $V-$ terminals through a reference resistance $R_0$, adjusted to the average conductance of the MTJ $(R_0^-1=GP+GAP/2)$ where $GP$ and $GAP$ represented conductance in parallel (P) and anti-parallel (AP) states, respectively. This voltage becomes an input to the CMOS inverters that are biased at the middle point of their DC operating point, creating a stochastic output whose probability can be tuned by the input charge current.  

\begin{table}[]
\centering
\small
\caption{Parameters for p-bit Based Activation Function.}
\label{tab:parameter}
\begin{tabular}{ccc}
\hline
Parameter & Description & Value  \\ \hline
\multicolumn{3}{l}{\textbf{Circular FM}} \\ \hline
$\phi$ & Diameter & $100 nm$ \\
$t$  & Thickness  & $2 nm$   \\
$\alpha$ & Damping coefficient & $0.01$  \\ \hline
\multicolumn{3}{l}{\textbf{MTJ}} \\ \hline
$G0$ & Conductance & $150e^{-6} S$ \\
$P$  & Spin Polarization & $0.52$ \\ \hline
\multicolumn{3}{l}{\textbf{\begin{tabular}[c]{@{}l@{}}Giant Spin Hall Layer(GSHE)\end{tabular}}}      \\ \hline
$\lambda$ & Spin-diffusion length & $2.1 nm$ \\
$\theta$      & Spin Hall Angle  & $0.5$ \\
$Volume$   & $l \times w \times t$ & $100 nm \times 100 nm \times 3.15 nm$ \\ \hline
\end{tabular}
\vspace{-0.3cm}
\end{table}

\begin{figure*}
\includegraphics[height=2.2in, width=5.2in]{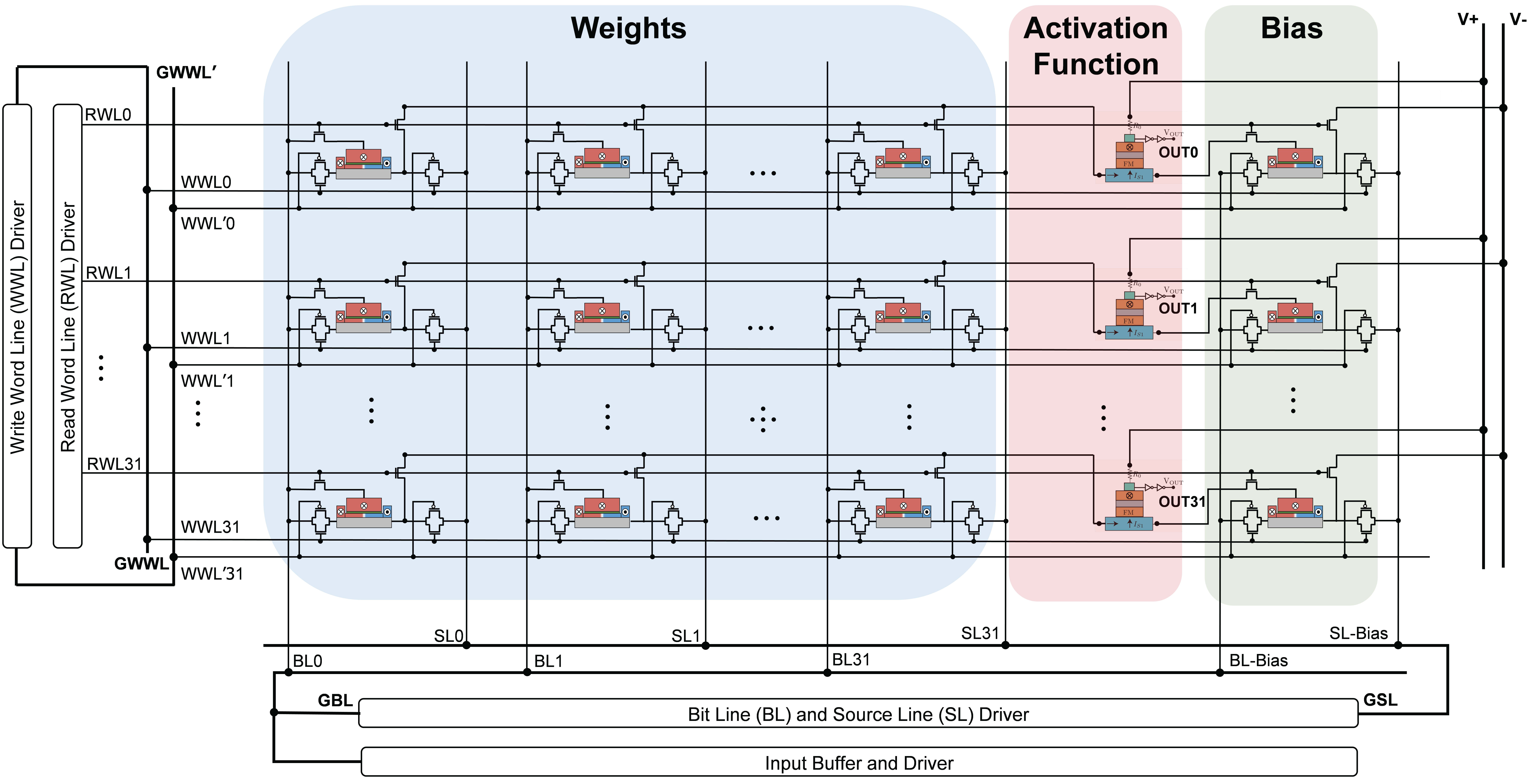}
\caption{Proposed $32 \times 32$ hybrid CMOS/spin-based weighted array structure for RBM implementation.}
\end{figure*}

Each component of the device is represented by an independent spin-circuit based on experimentally-benchmarked models that have been established in \cite{camsari2015modular} and simulated as a spin-circuit in a SPICE-like platform. Here, we obtain an analytical approximation to the time-averaged behavior of the output characteristics. We start by relating the charge current flowing in the spin Hall layer to the spin-current absorbed by the magnet, assuming short-circuit conditions for simplicity, i.e. 100\% spin absorption by the FM: 

\begin{equation}
  I_s/I_c = \beta = \frac{L}{t} (\theta) (1 - sech(\frac{t}{\lambda}))
\end{equation}
where $I_s$ is the spin-current, $I_c$ is the charge current, $\theta$ is the spin-Hall angle, $L$, $t$, $\lambda$ are the length, thickness and spin diffusion lengths for the spin-Hall layer. The length and width of the GSHE layer are assumed to be the same as the circular nanomagnet. With a suitable choice of the L and t, the spin-current generated can be greater in magnitude than the charge current generating ``gain.'' For the parameters used in this paper, which are listed in Table~\ref{tab:parameter}, the gain factor $\beta$ is $\sim 10$. Next, we approximate the behavior of the magnetization as a function of an input spin-current, polarized in the ($\pm z$) direction. For a magnet with only a PMA in the $\pm z$ direction, a distribution function at steady state can be written analytically as below, as long as the spin-current is also fully in the $\pm z$ direction:

\begin{equation}
  \rho (m_z) = \frac{1}{Z} exp(\Delta m_z^2 + 2 i_s m_z)
\end{equation}
where $Z$ is a normalization constant, $m_z$ is the magnetization along $+z$,  is the thermal barrier of the nanomagnet, and $i_s$ is a normalization quantity for the spin-current such that $i_s= I_s/(4q/ \hbar \alpha kT)$, $\alpha$ being the damping coefficient of the magnet, $q$ the electron charge and $\hbar$ the reduced Planck constant. It is possible to use (4) to obtain an average magnetization $<m_z> = \int_{-1}^{+1} d m_z m_z \rho (m_z) / \int_{-1}^{+1} d m_z \rho (m_z)$. Assuming $\Delta \ll kT$, $<m_z>$ can be evaluated to give the Langevin function, $<m_z> = L(i_s)$ where $L(x) = \frac{1}{x} - coth \frac{1}{x}$, which is an exact description for the average magnetization in the presence of a z-directed spin-current for a low barrier PMA magnet.

In the present case, however, the nanomagnet has a circular shape with a strong in-plane anisotropy and no simple analytical formula can be derived, thus We use the Langevin function with a fitting parameter that adjusts the normalization current by a factor $\eta$, so that the modified normalization constant becomes $(4 q/\hbar \alpha kT)(\eta)$. This factor increases with elevating the shape anisotropy $(H_d \sim 4 \pi M_s)$ and becomes exactly one when there is no shape anisotropy. Once the magnetization and charge currents are related, we can approximate the output probability of the CMOS inverters by a phenomenological equation along with fitting parameter $\chi$ as follows, $p = \frac{V_{OUT}}{VDD} \approx \frac{1}{2} [1-tanh(\chi <m_z>)]$,   
which allows us to relate the input charge current to the output probability, with physical parameters.  Fig. 2 shows the comparison of the full SPICE-model with respect to aforementioned equations showing good agreement with two fitting parameters $\eta$ and $\chi$, which fit the magnetization and CMOS components, respectively.

\section{Proposed Weighted Array Design}
Figure 3 shows the structure of the weighted array proposed herein to implement the RBM architecture including the SOT-DWM based weighted connections and biases, as well as the p-bit based activation functions. Transmission gates (TGs) are utilized in write circuits within the bit cells of the weighted connection to adjust weights by moving the DW position. As investigated in \cite{zand2017}, TGs can provide energy-efficient and symmetric switching operation for SOT-based devices, which are desirable during the training phase. Table~\ref{tab:signaling} lists the required signaling for controlling the training and read operations in the weighted array structure. Herein, a chain of inverters are considered to drive signal lines, in which each successive inverter is twice as large as the previous one.

During the read operation, write word line (WWL) is connected to ground (GND) and the source line (SL) is in high impedance (Hi-Z) state, which disconnects the write path. The read word line (RWL) for each row is connected to VDD, which turns ON the read transistors in the weighted connection bit cell. The bit line (BL) will be connected to the input signal (VIN), which results in producing a current that affects the output probability of the p-bit device. The direction of the generated current relies on the VIN signal. In particular, since V- is supplied by a voltage source equal to VDD/2, if VIN is connected to VDD the injected current to the p-bit based activation function will have positive value, and if VIN is zero the input current will be negative. The amplitude of the generated current depends on the resistance of the weighted connection which is defined by the position of the DW in the SOT-DWM device. 

During the training operation, the RWL is connected to GND, which turns OFF the read transistors and disconnects the read path. The WWL is connected to an input pulse (VPULSE) signal which activates the write path for a short period of time. The duration of the VPULSE should be designed in a manner such that it can provide the desired learning rate, $\eta$, to the training circuit. For instance, a high VPULSE duration results in a significant change in the DW position in each training iteration, which effectively reduces the number of different resistive states that can be realized by the SOT-DWM device. Resistance of the weighted connections can be adjusted by the BL and SL signals, as listed in Table~\ref{tab:signaling}. A higher resistance leads to a smaller current injected to the p-bit device. Therefore, the input signal connected to the weighted connection will have lower impact on the output probability of the p-bit device, which means the input signal exhibits a lower weight. The bias nodes can also be adjusted similar to the weighted connection.  

\begin{table}[]
\centering
\small
\caption{Signaling to Control The Array Operations.}
\vspace{-0.2cm}
\label{tab:signaling}
\begin{tabular}{ccccccc}
\hline
Operation & WWL & RWL & BL  & SL   & V+   & V-    \\ \hline
\begin{tabular}[c]{@{}c@{}}Increase Weight\end{tabular} & VPULSE & GND & VDD & GND  & Hi-Z & Hi-Z  \\
\begin{tabular}[c]{@{}c@{}}Decrease Weight\end{tabular} & VPULSE & GND & GND & GND  & Hi-Z & Hi-Z  \\
Read & GND    & VDD & VIN & Hi-Z & VDD  & VDD/2 \\ \hline
\end{tabular}
\vspace{-0.3cm}
\end{table}

\begin{table}[]
\centering
\small
\caption{Relation between the input currents of activation functions and array size for $R_P = 1 M \Omega$.}
\vspace{-0.2cm}
\label{tab:arraysize}
\begin{tabular}{ccccc}
\hline
\multirow{2}{*}{Features}       & \multicolumn{4}{c}{Array Size}                                  \\ \cline{2-5} 
                                & $8 \times 8$ & $16 \times 16$ & $32 \times 32$ & $64 \times 64$ \\ \hline
Max. Positive Current $(\mu A)$ & 2.71         & 5.14           & 10.79          & 21.46          \\
Max. Negative Current $(\mu A)$ & 3.57         & 7.14           & 14.23          & 28.28          \\
Max. output ``0'' Probability   & 0.77         & 0.88           & 0.95           & 0.97           \\
Min. output ``0'' Probability   & 0.175        & 0.08           & 0.038          & 0.026          \\ \hline
\end{tabular}
\end{table}

\section{Simulation Results And Discussion}
To analyze the RBM implementation using the proposed p-bit device and the weighted array structure, we have utilized a hierarchical simulation framework including circuit-level and application-level simulations. In circuit level simulation, the behavioral models of the p-bit and SOT-DWM devices were leveraged in SPICE circuit simulations using 20nm CMOS technology with 0.9V nominal voltage to validate the functionality of the designed weighted array circuit. In application-level simulation, the results obtained from device-level and circuit-level simulations are used to implement a DBN architecture and analyze its behavior in MATLAB.

\begin{figure}
\includegraphics[scale=0.22]{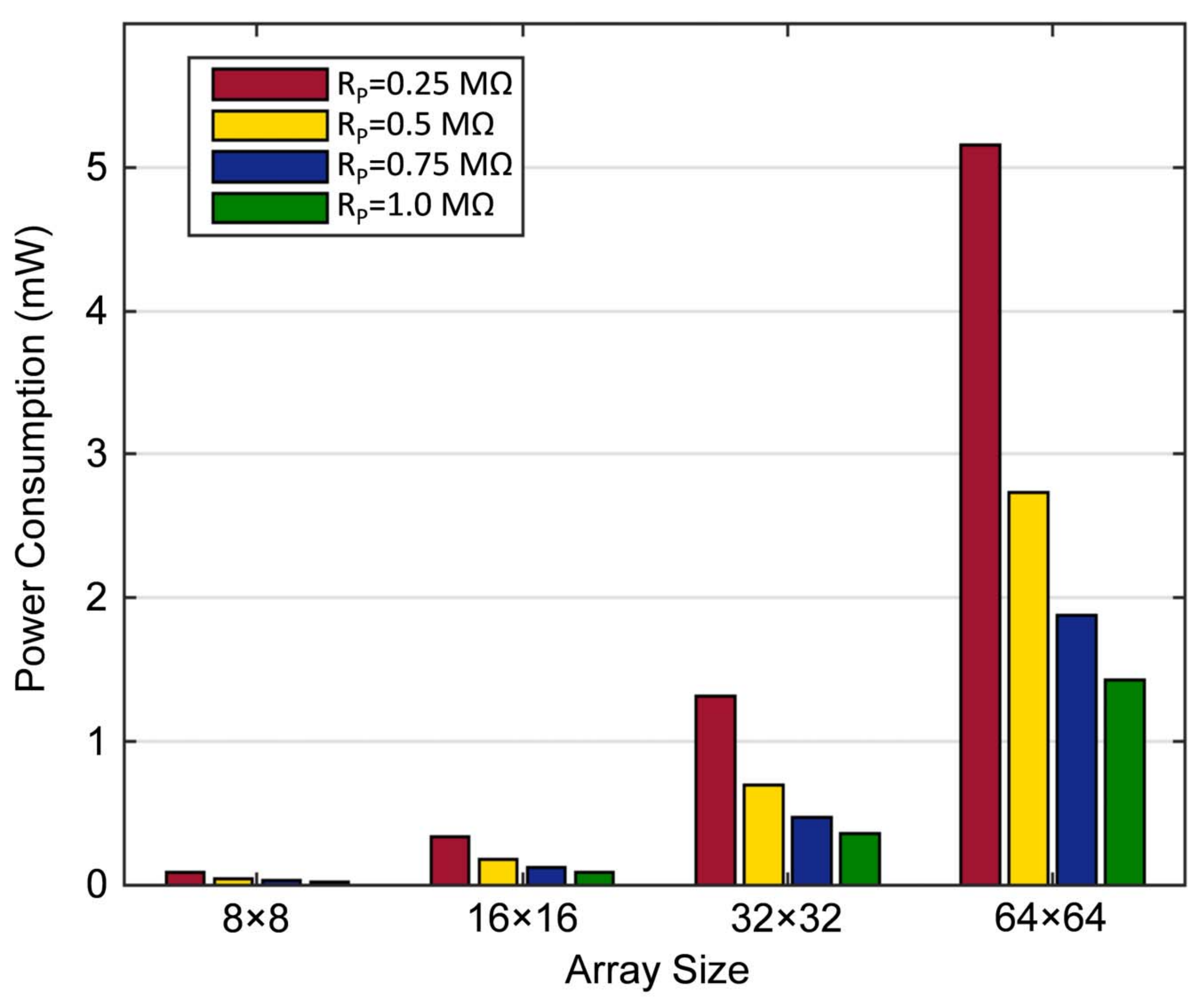}
\caption{Weighted array power consumption versus the resistance of the weighted connections and array size.}
\vspace{-0.3cm}
\end{figure}

\subsection{Circuit-level simulation}
The device-level simulations shown in Fig. 2 verified a sigmoidal relation between the input current of the p-bit based activation function and its output probability. The shape of the activation on function is one of the major factors affecting the performance of the RBM. Therefore, we have provided comprehensive analyses on the impacts of weighted connection resistance and weighted array dimensions on the input currents of the p-bit based activation functions, and the power consumption of the weighted array.

Table~\ref{tab:arraysize} lists the range of the activation function input currents for various weighted array dimensions, while the resistance of the SOT-DWM device in parallel state (RP) is constant and equals $1 M\Omega$.  The experimental results provided in [19, 28] exhibit that an MTJ resistance in the $M\Omega$ range can be obtained by increasing the oxide thickness in an MTJ structure. The highest positive and negative currents can be achieved while the weighted connections are in parallel state, i.e. lowest resistance, and all of the input voltages (VIN) are equal to VDD and GND, respectively. The difference between the amplitude of positive and negative currents in a given array size with constant RP is caused by the different pull-down and pull-up strengths in NMOS read transistors. The maximum and minimum output-level ``0'' probabilities are listed in Table~\ref{tab:arraysize}, which can be obtained according to the measured input currents and the sigmoidal activation function shown in Fig. 2.

\begin{table}[]
\centering
\small
\caption{Relation between the input currents of activation functions and $R_P$ in a $32 \times 32$ array.}
\vspace{-0.2cm}
\label{tab:rp}
\begin{tabular}{ccccc}
\hline
\multirow{2}{*}{Features}       & \multicolumn{4}{c}{$R_P (M \Omega)$} \\ \cline{2-5} 
                                & 0.25    & 0.5     & 0.75    & 1      \\ \hline
Max. Positive Current $(\mu A)$ & 36.56   & 20.02   & 13.97   & 10.79  \\
Max. Negative Current $(\mu A)$ & 54.95   & 28.12   & 18.9    & 14.23  \\
Max. output ``0'' Probability   & 0.98    & 0.965   & 0.96    & 0.95   \\
Min. output ``0'' Probability   & 0.01    & 0.026   & 0.032   & 0.038  \\ \hline
\vspace{-0.4cm}
\end{tabular}
\end{table}

\begin{table*}[]
\centering
\small
\caption{Comparison between various RBM implementations with an emphasis on activation function structure.} 
\label{tab:compare}
\begin{tabular}{cccccccc} \hline

\begin{tabular}[c]{@{}c@{}}Design\end{tabular}                    & {\cite{Kim2010}}  & {\cite{Ly2010}} & {\cite{Ardakani2017}} & {\cite{SHERI2015}} & {\cite{Bojnordi2016}} & {\cite{Eryilmaz2016}} & \begin{tabular}[c]{@{}c@{}}Proposed\\ Herein\end{tabular}\\  \hline

\begin{tabular}[c]{@{}c@{}}Weighted\\ Connection\end{tabular}        & \begin{tabular}[c]{@{}c@{}}Embedded\\ multipliers \end{tabular}  & \begin{tabular}[c]{@{}c@{}}Embedded\\ multipliers \end{tabular} & \begin{tabular}[c]{@{}c@{}}- LFSR\\  - AND/OR gates\end{tabular} &  \begin{tabular}[c]{@{}c@{}}RRAM\\ memristor \end{tabular} & RRAM & PCM & SOT-DWM  \\ \hline

\begin{tabular}[c]{@{}c@{}}Activation\\ Function\end{tabular}        & \begin{tabular}[c]{@{}c@{}}CMOS-based \\ LUTs\end{tabular} & \begin{tabular}[c]{@{}c@{}}-2-kB BRAM\\ - Picewise Linear \\Interpolator\\ - Random number\\ Generator\end{tabular} & \begin{tabular}[c]{@{}c@{}} - LFSR\\ - Bit-wise AND\\ - tree adder\\ - FSM-based \\tanh unit\end{tabular} & Off-chip & \begin{tabular}[c]{@{}c@{}} - $64 \times 16$ LUTs\\ - Pseudo Random \\Number Generator\\ - Comparator\end{tabular} & Off-chip                 & \begin{tabular}[c]{@{}c@{}} - near-zero \\energy barrier \\probabilistic \\spin logic \\ device\end{tabular} \\ \hline
\begin{tabular}[c]{@{}c@{}}Energy per neuron\end{tabular}          & N/A & $\sim 10-100 nJ$ & $\sim 10-100 pJ$ & N/A & $\sim 1-10 nJ$ & N/A    & $\sim 1-10 fJ$  \\ \hline
\begin{tabular}[c]{@{}c@{}}Normalized area per neuron\end{tabular} & N/A  & $\sim 3000 \times$ & $\sim 90 \times$ & N/A & $\sim 1250 \times$ & N/A    & $\sim 1 \times$ \\                                                               
\hline
\end{tabular} 
                
\end{table*}

Moreover, Table~\ref{tab:rp} illustrates the relation between the $R_P$ values and input currents of the activation functions, and their corresponding output probabilities, for a given $32 \times 32$ weighted array. The lower RP resistance and higher array size provides a wider range of output probabilities which can increase the RBM performance. However, this is achieved at the cost of higher area and power consumption. The trade-offs between the array size, weighted connection resistance, and average power consumption in a single read operation is shown in Fig. 4. The lowest power consumption of 22.6 $\mu W$ is realized by an $8 \times 8$ array with $R_P = 1 M \Omega$. However, this array provides the narrowest range of the output probabilities, which significantly reduces the performance of the DBN.

\subsection{Application-level simulation}
In the application-level simulation, we have leveraged the obtained device and circuit behavioral models to simulate a DBN architecture for digit recognition. In particular, learning rate and the shape of the sigmoid activation function is extracted by the SOT-DWM and p-bit device-level simulations, respectively, while the circuit-level simulations defines the range of the output probabilities.  To evaluate the performance of the system, we have modified a MATLAB implementation of DBN by Tanaka and Okutomi \cite{Tanaka2014} and used the MNIST data set \cite{Lecun1998} including 60,000 and 10,000 sample images with $28 \times 28$ pixels for training and testing operations, respectively. We have used Error rate (ERR) metric to evaluate the performance of the DBN, as expressed by $ERR=N_F/N$, where, $N$ is the number of input data, $N_F$ is the number of false inference \cite{Tanaka2014}.

The simplest model of the DBN that can be implemented for MNIST digit recognition consists 784 nodes in visible layer to handle $28 \times 28$ pixels of the input images, and 10 nodes in hidden layer representing the output classes. Fig. 5 shows the relation between the performance of various DBN topologies, and the number of input training samples ranging from 100 to 5,000, which is obtained using 1,000 test samples. The ERR and RMSE metrics can be improved by enlarging the DBN structure through increasing the number of hidden layers, as well as the number of nodes in each layer. This improvement is realized at the cost of larger area and power consumptions. Increasing the input training samples can improve the DBN performance as well, however it will quickly converge due to the limited weight values that can be provided by SOT-DWM based weighted connections. As shown in Fig. 5, some random behaviors are observed for networks with smaller sizes that are trained by lower number of training samples, which will be significantly reduced by increasing the number of training samples.

The simulation results exhibit the highest error rate of 36.8\% for a $784 \times 10$ DBN that is trained by 100 training samples. Meanwhile, the lowest error rate of 3.7\% was achieved using a $784 \times 800 \times 800 \times 10$ DBN trained by 5,000 input training samples. This illustrates that the recognition error rate can be decreased by increasing the number of hidden layers, and training samples, which is also realized at the cost of higher area and power overheads.

\begin{figure}
\includegraphics[scale=0.5]{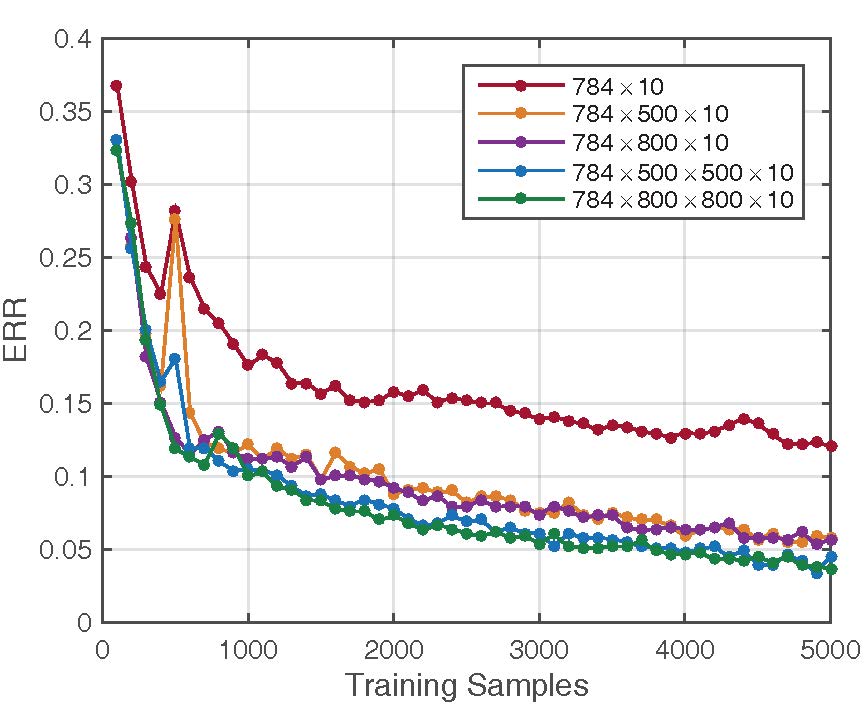}
\vspace{-0.2cm}
\caption{ERR for various DBN topologies.}
\vspace{-0.4cm}
\end{figure}

\subsection{Disucussion}
Table~\ref{tab:compare} lists previous hardware-based RBM implementations, which have aimed to overcome software limitations by utilizing FPGAs \cite{Kim2010,Ly2010}, stochastic CMOS \cite{Ardakani2017}, and hybrid memristor-CMOS designs \cite{SHERI2015,Bojnordi2016,Eryilmaz2016}. FPGA implementations demonstrated RBM speedups of 25-145 over software implementations \cite{Kim2010,Ly2010}, but had significant constraints such as only realizing a single $128 \times 128$ RBM per FPGA chip, routing congestion, and clock frequencies limited to 100MHz \cite{Ly2010}. The stochastic CMOS-based RBM implementation proposed in \cite{Ardakani2017} leveraged the low-complexity of stochastic CMOS arithmetic to save area and power. However, the need for extremely long bit-stream lengths negate energy savings and lead to very long latencies. Additionally, a significant amount of Linear Feedback Shift Registers (LFSRs) were required to produce the uncorrelated input and weight bit-streams. In both the FPGA and stochastic CMOS designs, improvements were achieved by implementing parallel Boolean circuits such as multipliers and pseudo-random number generators for probabilistic behavior, which has significant area and energy overheads compared to leveraging the physical behaviors of emerging devices to perform the computation intrinsically. Bojnordi et al. \cite{Bojnordi2016} leveraged resistive RAM (RRAM) devices to implement efficient matrix multiplication for weighted products within Boltzmann machine applications, and demonstrated significant speedup of up to 100-fold over single-threaded cores and energy savings of over 10-fold. Similarly, Sheri et al. \cite{SHERI2015} and Eryilmaz et al. \cite{Eryilmaz2016} utilized RRAM and PCM devices to implement matrix multiplication, while the corresponding activation function circuitry is still based on the CMOS technology, which suffers from the aforementioned area and power consumption overheads.

While most of the previous hybrid Memristor/CMOS designs focus on improving the performance of weighted connections, the work presented herein overcomes many of the preceding challenges of generating sigmoidal probabilistic activation functions by utilizing a novel p-bit device that leverages intrinsic thermal noise within low energy barrier nanomagnets to provide a natural building block for RBMs within a compact and low-energy package. As listed in Table V, the proposed design can achieve approximately three orders of magnitude improvement in term of energy consumption compared to the most energy-efficient designs, while realizing at least ~90X device count reduction for considerable area savings. Note that these calculations do not take into account the weighted connections, since the main focus of this paper is on the activation function. While SOT-DWM devices are utilized herein for the weighted connections, any other memristive devices could be utilized without loss of generality.  

\section{Conclusion}
Herein, we developed a hybrid CMOS/spin-based DBN implementation using p-bit based activation functions modeled to produce a probabilistic output that can be modulated by an input current. The device-level simulations exhibited a sigmoid relation between the input currents and output probability. The SPICE model of the p-bit is used to design a weighted array structure to implement RBM. The circuit simulations showed that the performance of the array can be improved by enlarging the array size, as well as reducing the resistance of the weighted connections. However, these improvements are achieved at the cost of increased area and power consumption. For instance, the lowest power dissipation among the examined designs belongs to an $8 \times 8$ array with the maximum resistance of $1 M \Omega$ for weighted connections. However, this structure can only provide the output probabilities ranging from 0.175 to 0.77, which is the narrowest range among the examined designs resulting in a DBN implementation with lowest accuracy. 

Next, we simulated a DBN for digit recognition application in MATLAB using the device and circuit-level behavioral models. Trade-offs include the relations between the recognition accuracy of the DBN and the number of training samples, which are comparable to conventional hardware implementations. The recognition error rate decreased substantially for the first thousand training samples, regardless of the size of the array, while benefits continue through several thousand inputs. However, at least two hidden layers are desirable to achieve suitable error rates. Finally, we have provided a comparison between previous hardware-based RBM implementations and our design with an emphasis on the probabilistic activation function within the neuron structure. The results exhibited that the p-bit based activation function can achieve roughly three orders of magnitude energy improvement, while realizing at least 90X reduction in terms of device count, compared to the previous most energy-efficient designs. The research directions herein enable several intriguing possibilities for future work, including: (1) implementing the entire network in SPICE to obtain more robust results; 2) investigating the effect of process variation and noise on the accuracy of proposed architecture; 3) studying alternative devices with lower susceptibility to thermal noise; and 4) studying the scalability challenges of DBNs using larger datasets, e.g. CIFAR.

\section{Acknowledgements}
This work was supported in part by the Center for Probabilistic Spin Logic for Low-Energy Boolean and Non-Boolean Computing  (CAPSL), one of the Nanoelectronic Computing Research (nCORE) Centers as task 2759.006, a Semiconductor Research Corporation (SRC) program sponsored by the NSF through CCF 1739635.

\bibliographystyle{ACM-Reference-Format}
\bibliography{sample-bibliography} 


\begin{thebibliography}{24}


\ifx \showCODEN    \undefined \def \showCODEN     #1{\unskip}     \fi
\ifx \showDOI      \undefined \def \showDOI       #1{#1}\fi
\ifx \showISBNx    \undefined \def \showISBNx     #1{\unskip}     \fi
\ifx \showISBNxiii \undefined \def \showISBNxiii  #1{\unskip}     \fi
\ifx \showISSN     \undefined \def \showISSN      #1{\unskip}     \fi
\ifx \showLCCN     \undefined \def \showLCCN      #1{\unskip}     \fi
\ifx \shownote     \undefined \def \shownote      #1{#1}          \fi
\ifx \showarticletitle \undefined \def \showarticletitle #1{#1}   \fi
\ifx \showURL      \undefined \def \showURL       {\relax}        \fi
\providecommand\bibfield[2]{#2}
\providecommand\bibinfo[2]{#2}
\providecommand\natexlab[1]{#1}
\providecommand\showeprint[2][]{arXiv:#2}

\bibitem[\protect\citeauthoryear{Ackley, Hinton, and Sejnowski}{Ackley
  et~al\mbox{.}}{1985}]%
        {ackley1985}
\bibfield{author}{\bibinfo{person}{David~H Ackley}, \bibinfo{person}{Geoffrey~E
  Hinton}, {and} \bibinfo{person}{Terrence~J Sejnowski}.}
  \bibinfo{year}{1985}\natexlab{}.
\newblock \showarticletitle{A learning algorithm for Boltzmann machines}.
\newblock \bibinfo{journal}{{\em Cognitive science\/}} \bibinfo{volume}{9},
  \bibinfo{number}{1} (\bibinfo{year}{1985}), \bibinfo{pages}{147--169}.
\newblock


\bibitem[\protect\citeauthoryear{Ardakani, Leduc-Primeau, Onizawa, Hanyu, and
  Gross}{Ardakani et~al\mbox{.}}{2017}]%
        {Ardakani2017}
\bibfield{author}{\bibinfo{person}{A. Ardakani}, \bibinfo{person}{F.
  Leduc-Primeau}, \bibinfo{person}{N. Onizawa}, \bibinfo{person}{T. Hanyu},
  {and} \bibinfo{person}{W.~J. Gross}.} \bibinfo{year}{2017}\natexlab{}.
\newblock \showarticletitle{VLSI Implementation of Deep Neural Network Using
  Integral Stochastic Computing}.
\newblock \bibinfo{journal}{{\em IEEE Transactions on Very Large Scale
  Integration (VLSI) Systems\/}} \bibinfo{volume}{25}, \bibinfo{number}{10}
  (\bibinfo{year}{2017}).
\newblock
\showISSN{1063-8210}


\bibitem[\protect\citeauthoryear{Behin-Aein, Diep, and Datta}{Behin-Aein
  et~al\mbox{.}}{2016}]%
        {behin2016}
\bibfield{author}{\bibinfo{person}{Behtash Behin-Aein}, \bibinfo{person}{Vinh
  Diep}, {and} \bibinfo{person}{Supriyo Datta}.}
  \bibinfo{year}{2016}\natexlab{}.
\newblock \showarticletitle{A building block for hardware belief networks}.
\newblock \bibinfo{journal}{{\em Scientific reports\/}}  \bibinfo{volume}{6}
  (\bibinfo{year}{2016}).
\newblock


\bibitem[\protect\citeauthoryear{Bojnordi and Ipek}{Bojnordi and Ipek}{2016}]%
        {Bojnordi2016}
\bibfield{author}{\bibinfo{person}{M.~N. Bojnordi} {and} \bibinfo{person}{E.
  Ipek}.} \bibinfo{year}{2016}\natexlab{}.
\newblock \showarticletitle{Memristive Boltzmann machine: A hardware
  accelerator for combinatorial optimization and deep learning}. In
  \bibinfo{booktitle}{{\em 2016 IEEE International Symposium on High
  Performance Computer Architecture (HPCA)}}.
\newblock


\bibitem[\protect\citeauthoryear{Buesing, Bill, Nessler, and Maass}{Buesing
  et~al\mbox{.}}{2011}]%
        {buesing2011}
\bibfield{author}{\bibinfo{person}{Lars Buesing}, \bibinfo{person}{Johannes
  Bill}, \bibinfo{person}{Bernhard Nessler}, {and} \bibinfo{person}{Wolfgang
  Maass}.} \bibinfo{year}{2011}\natexlab{}.
\newblock \showarticletitle{Neural dynamics as sampling: a model for stochastic
  computation in recurrent networks of spiking neurons}.
\newblock \bibinfo{journal}{{\em PLoS computational biology\/}}
  \bibinfo{volume}{7}, \bibinfo{number}{11} (\bibinfo{year}{2011}),
  \bibinfo{pages}{e1002211}.
\newblock


\bibitem[\protect\citeauthoryear{Camsari, Faria, Sutton, and Datta}{Camsari
  et~al\mbox{.}}{2017}]%
        {Camsari2017}
\bibfield{author}{\bibinfo{person}{Kerem~Yunus Camsari},
  \bibinfo{person}{Rafatul Faria}, \bibinfo{person}{Brian~M. Sutton}, {and}
  \bibinfo{person}{Supriyo Datta}.} \bibinfo{year}{2017}\natexlab{}.
\newblock \showarticletitle{Stochastic $p$-Bits for Invertible Logic}.
\newblock \bibinfo{journal}{{\em Phys. Rev. X\/}}  \bibinfo{volume}{7}
  (\bibinfo{date}{Jul} \bibinfo{year}{2017}), \bibinfo{pages}{031014}.
\newblock
Issue 3.


\bibitem[\protect\citeauthoryear{Camsari, Ganguly, and Datta}{Camsari
  et~al\mbox{.}}{2015}]%
        {camsari2015modular}
\bibfield{author}{\bibinfo{person}{Kerem~Yunus Camsari},
  \bibinfo{person}{Samiran Ganguly}, {and} \bibinfo{person}{Supriyo Datta}.}
  \bibinfo{year}{2015}\natexlab{}.
\newblock \showarticletitle{Modular approach to spintronics}.
\newblock \bibinfo{journal}{{\em Scientific reports\/}}  \bibinfo{volume}{5}
  (\bibinfo{year}{2015}).
\newblock


\bibitem[\protect\citeauthoryear{Carreira-Perpinan and
  Hinton}{Carreira-Perpinan and Hinton}{2005}]%
        {carreira2005}
\bibfield{author}{\bibinfo{person}{Miguel~A Carreira-Perpinan} {and}
  \bibinfo{person}{Geoffrey~E Hinton}.} \bibinfo{year}{2005}\natexlab{}.
\newblock \showarticletitle{On contrastive divergence learning.}. In
  \bibinfo{booktitle}{{\em Aistats}}, Vol.~\bibinfo{volume}{10}.
  \bibinfo{pages}{33--40}.
\newblock


\bibitem[\protect\citeauthoryear{Eryilmaz, Neftci, Joshi, Kim, BrightSky, Lung,
  Lam, Cauwenberghs, and Wong}{Eryilmaz et~al\mbox{.}}{2016}]%
        {Eryilmaz2016}
\bibfield{author}{\bibinfo{person}{S.~B. Eryilmaz}, \bibinfo{person}{E.
  Neftci}, \bibinfo{person}{S. Joshi}, \bibinfo{person}{S. Kim},
  \bibinfo{person}{M. BrightSky}, \bibinfo{person}{H.~L. Lung},
  \bibinfo{person}{C. Lam}, \bibinfo{person}{G. Cauwenberghs}, {and}
  \bibinfo{person}{H.~S.~P. Wong}.} \bibinfo{year}{2016}\natexlab{}.
\newblock \showarticletitle{Training a Probabilistic Graphical Model With
  Resistive Switching Electronic Synapses}.
\newblock \bibinfo{journal}{{\em IEEE Transactions on Electron Devices\/}}
  \bibinfo{volume}{63}, \bibinfo{number}{12} (\bibinfo{date}{Dec}
  \bibinfo{year}{2016}), \bibinfo{pages}{5004--5011}.
\newblock
\showISSN{0018-9383}


\bibitem[\protect\citeauthoryear{Faria, Camsari, and Datta}{Faria
  et~al\mbox{.}}{2017}]%
        {Faria2017}
\bibfield{author}{\bibinfo{person}{R. Faria}, \bibinfo{person}{K.~Y. Camsari},
  {and} \bibinfo{person}{S. Datta}.} \bibinfo{year}{2017}\natexlab{}.
\newblock \showarticletitle{Low-Barrier Nanomagnets as p-Bits for Spin Logic}.
\newblock \bibinfo{journal}{{\em IEEE Magnetics Letters\/}}
  \bibinfo{volume}{8} (\bibinfo{year}{2017}), \bibinfo{pages}{1--5}.
\newblock
\showISSN{1949-307X}


\bibitem[\protect\citeauthoryear{Hinton, Osindero, and Teh}{Hinton
  et~al\mbox{.}}{2006}]%
        {hinton2006}
\bibfield{author}{\bibinfo{person}{Geoffrey~E Hinton}, \bibinfo{person}{Simon
  Osindero}, {and} \bibinfo{person}{Yee-Whye Teh}.}
  \bibinfo{year}{2006}\natexlab{}.
\newblock \showarticletitle{A fast learning algorithm for deep belief nets}.
\newblock \bibinfo{journal}{{\em Neural computation\/}} \bibinfo{volume}{18},
  \bibinfo{number}{7} (\bibinfo{year}{2006}), \bibinfo{pages}{1527--1554}.
\newblock


\bibitem[\protect\citeauthoryear{Kim, McMahon, and Olukotun}{Kim
  et~al\mbox{.}}{2010}]%
        {Kim2010}
\bibfield{author}{\bibinfo{person}{S.~K. Kim}, \bibinfo{person}{P.~L. McMahon},
  {and} \bibinfo{person}{K. Olukotun}.} \bibinfo{year}{2010}\natexlab{}.
\newblock \showarticletitle{A Large-Scale Architecture for Restricted Boltzmann
  Machines}. In \bibinfo{booktitle}{{\em 2010 18th IEEE Annual International
  Symposium on Field-Programmable Custom Computing Machines}}.
  \bibinfo{pages}{201--208}.
\newblock


\bibitem[\protect\citeauthoryear{Lecun, Bottou, Bengio, and Haffner}{Lecun
  et~al\mbox{.}}{1998}]%
        {Lecun1998}
\bibfield{author}{\bibinfo{person}{Y. Lecun}, \bibinfo{person}{L. Bottou},
  \bibinfo{person}{Y. Bengio}, {and} \bibinfo{person}{P. Haffner}.}
  \bibinfo{year}{1998}\natexlab{}.
\newblock \showarticletitle{Gradient-based learning applied to document
  recognition}.
\newblock \bibinfo{journal}{{\it Proc. IEEE}} \bibinfo{volume}{86},
  \bibinfo{number}{11} (\bibinfo{date}{Nov} \bibinfo{year}{1998}),
  \bibinfo{pages}{2278--2324}.
\newblock
\showISSN{0018-9219}


\bibitem[\protect\citeauthoryear{Liu, Pai, Li, Tseng, Ralph, and Buhrman}{Liu
  et~al\mbox{.}}{2012}]%
        {Liu2012}
\bibfield{author}{\bibinfo{person}{Luqiao Liu}, \bibinfo{person}{Chi-Feng Pai},
  \bibinfo{person}{Y. Li}, \bibinfo{person}{H.~W. Tseng},
  \bibinfo{person}{D.~C. Ralph}, {and} \bibinfo{person}{R.~A. Buhrman}.}
  \bibinfo{year}{2012}\natexlab{}.
\newblock \showarticletitle{Spin-Torque Switching with the Giant Spin Hall
  Effect of Tantalum}.
\newblock \bibinfo{journal}{{\em Science\/}} \bibinfo{volume}{336},
  \bibinfo{number}{6081} (\bibinfo{year}{2012}), \bibinfo{pages}{555--558}.
\newblock
\showISSN{0036-8075}


\bibitem[\protect\citeauthoryear{Ly and Chow}{Ly and Chow}{2010}]%
        {Ly2010}
\bibfield{author}{\bibinfo{person}{D.~L. Ly} {and} \bibinfo{person}{P. Chow}.}
  \bibinfo{year}{2010}\natexlab{}.
\newblock \showarticletitle{High-Performance Reconfigurable Hardware
  Architecture for Restricted Boltzmann Machines}.
\newblock \bibinfo{journal}{{\em IEEE Transactions on Neural Networks\/}}
  \bibinfo{volume}{21}, \bibinfo{number}{11} (\bibinfo{date}{Nov}
  \bibinfo{year}{2010}), \bibinfo{pages}{1780--1792}.
\newblock
\showISSN{1045-9227}


\bibitem[\protect\citeauthoryear{Merolla, Arthur, Alvarez-Icaza, Cassidy,
  Sawada, Akopyan, Jackson, Imam, Guo, Nakamura, et~al\mbox{.}}{Merolla
  et~al\mbox{.}}{2014}]%
        {Merolla2014}
\bibfield{author}{\bibinfo{person}{Paul~A Merolla}, \bibinfo{person}{John~V
  Arthur}, \bibinfo{person}{Rodrigo Alvarez-Icaza}, \bibinfo{person}{Andrew~S
  Cassidy}, \bibinfo{person}{Jun Sawada}, \bibinfo{person}{Filipp Akopyan},
  \bibinfo{person}{Bryan~L Jackson}, \bibinfo{person}{Nabil Imam},
  \bibinfo{person}{Chen Guo}, \bibinfo{person}{Yutaka Nakamura},
  {et~al\mbox{.}}} \bibinfo{year}{2014}\natexlab{}.
\newblock \showarticletitle{A million spiking-neuron integrated circuit with a
  scalable communication network and interface}.
\newblock \bibinfo{journal}{{\em Science\/}} \bibinfo{volume}{345},
  \bibinfo{number}{6197} (\bibinfo{year}{2014}), \bibinfo{pages}{668--673}.
\newblock


\bibitem[\protect\citeauthoryear{Sarikaya, Hinton, and Deoras}{Sarikaya
  et~al\mbox{.}}{2014}]%
        {Sarikaya2014}
\bibfield{author}{\bibinfo{person}{Ruhi Sarikaya}, \bibinfo{person}{Geoffrey~E.
  Hinton}, {and} \bibinfo{person}{Anoop Deoras}.}
  \bibinfo{year}{2014}\natexlab{}.
\newblock \showarticletitle{Application of Deep Belief Networks for Natural
  Language Understanding}.
\newblock \bibinfo{journal}{{\em IEEE/ACM Trans. Audio, Speech and Lang.
  Proc.\/}} \bibinfo{volume}{22}, \bibinfo{number}{4} (\bibinfo{date}{April}
  \bibinfo{year}{2014}), \bibinfo{pages}{778--784}.
\newblock
\showISSN{2329-9290}


\bibitem[\protect\citeauthoryear{Sengupta, Banerjee, and Roy}{Sengupta
  et~al\mbox{.}}{2016a}]%
        {Sengupta2016hybrid}
\bibfield{author}{\bibinfo{person}{Abhronil Sengupta},
  \bibinfo{person}{Aparajita Banerjee}, {and} \bibinfo{person}{Kaushik Roy}.}
  \bibinfo{year}{2016}\natexlab{a}.
\newblock \showarticletitle{Hybrid Spintronic-CMOS Spiking Neural Network with
  On-Chip Learning: Devices, Circuits, and Systems}.
\newblock \bibinfo{journal}{{\em Phys. Rev. Applied\/}}  \bibinfo{volume}{6}
  (\bibinfo{date}{Dec} \bibinfo{year}{2016}), \bibinfo{pages}{064003}.
\newblock
Issue 6.


\bibitem[\protect\citeauthoryear{Sengupta, Panda, Wijesinghe, Kim, and
  Roy}{Sengupta et~al\mbox{.}}{2016b}]%
        {sengupta2016magnetic}
\bibfield{author}{\bibinfo{person}{Abhronil Sengupta},
  \bibinfo{person}{Priyadarshini Panda}, \bibinfo{person}{Parami Wijesinghe},
  \bibinfo{person}{Yusung Kim}, {and} \bibinfo{person}{Kaushik Roy}.}
  \bibinfo{year}{2016}\natexlab{b}.
\newblock \showarticletitle{Magnetic tunnel junction mimics stochastic cortical
  spiking neurons}.
\newblock \bibinfo{journal}{{\em Scientific reports\/}}  \bibinfo{volume}{6}
  (\bibinfo{year}{2016}), \bibinfo{pages}{30039}.
\newblock


\bibitem[\protect\citeauthoryear{Sengupta, Parsa, Han, and Roy}{Sengupta
  et~al\mbox{.}}{2016c}]%
        {Sengupta2016prob}
\bibfield{author}{\bibinfo{person}{A. Sengupta}, \bibinfo{person}{M. Parsa},
  \bibinfo{person}{B. Han}, {and} \bibinfo{person}{K. Roy}.}
  \bibinfo{year}{2016}\natexlab{c}.
\newblock \showarticletitle{Probabilistic Deep Spiking Neural Systems Enabled
  by Magnetic Tunnel Junction}.
\newblock \bibinfo{journal}{{\em IEEE Transactions on Electron Devices\/}}
  \bibinfo{volume}{63}, \bibinfo{number}{7} (\bibinfo{date}{July}
  \bibinfo{year}{2016}), \bibinfo{pages}{2963--2970}.
\newblock
\showISSN{0018-9383}


\bibitem[\protect\citeauthoryear{Sheri, Rafique, Pedrycz, and Jeon}{Sheri
  et~al\mbox{.}}{2015}]%
        {SHERI2015}
\bibfield{author}{\bibinfo{person}{Ahmad~Muqeem Sheri}, \bibinfo{person}{Aasim
  Rafique}, \bibinfo{person}{Witold Pedrycz}, {and} \bibinfo{person}{Moongu
  Jeon}.} \bibinfo{year}{2015}\natexlab{}.
\newblock \showarticletitle{Contrastive divergence for memristor-based
  restricted Boltzmann machine}.
\newblock \bibinfo{journal}{{\em Engineering Applications of Artificial
  Intelligence\/}}  \bibinfo{volume}{37} (\bibinfo{year}{2015}),
  \bibinfo{pages}{336 -- 342}.
\newblock
\showISSN{0952-1976}


\bibitem[\protect\citeauthoryear{Sutton, Camsari, Behin-Aein, and Datta}{Sutton
  et~al\mbox{.}}{2017}]%
        {sutton2017}
\bibfield{author}{\bibinfo{person}{Brian Sutton}, \bibinfo{person}{Kerem~Yunus
  Camsari}, \bibinfo{person}{Behtash Behin-Aein}, {and}
  \bibinfo{person}{Supriyo Datta}.} \bibinfo{year}{2017}\natexlab{}.
\newblock \showarticletitle{Intrinsic optimization using stochastic
  nanomagnets}.
\newblock \bibinfo{journal}{{\em Scientific Reports\/}}  \bibinfo{volume}{7}
  (\bibinfo{year}{2017}).
\newblock


\bibitem[\protect\citeauthoryear{Tanaka and Okutomi}{Tanaka and
  Okutomi}{2014}]%
        {Tanaka2014}
\bibfield{author}{\bibinfo{person}{M. Tanaka} {and} \bibinfo{person}{M.
  Okutomi}.} \bibinfo{year}{2014}\natexlab{}.
\newblock \showarticletitle{A Novel Inference of a Restricted Boltzmann
  Machine}. In \bibinfo{booktitle}{{\em 2014 22nd International Conference on
  Pattern Recognition}}. \bibinfo{pages}{1526--1531}.
\newblock
\showISSN{1051-4651}


\bibitem[\protect\citeauthoryear{Zand, Roohi, and DeMara}{Zand
  et~al\mbox{.}}{2017}]%
        {zand2017}
\bibfield{author}{\bibinfo{person}{R. Zand}, \bibinfo{person}{A. Roohi}, {and}
  \bibinfo{person}{R.~F. DeMara}.} \bibinfo{year}{2017}\natexlab{}.
\newblock \showarticletitle{Energy-Efficient and Process-Variation-Resilient
  Write Circuit Schemes for Spin Hall Effect MRAM Device}.
\newblock \bibinfo{journal}{{\em IEEE Transactions on Very Large Scale
  Integration (VLSI) Systems\/}} \bibinfo{volume}{25}, \bibinfo{number}{9}
  (\bibinfo{year}{2017}), \bibinfo{pages}{2394--2401}.
\newblock
\showISSN{1063-8210}


\end{thebibliography}

\end{document}